\documentclass[aps,prl,twocolumn,superscriptaddress]{revtex4-1}

\usepackage{graphicx}
\usepackage{amsmath}
\usepackage{braket}
\usepackage{verbatim}
\usepackage{color}
\usepackage{natbib}

\begin{document}

\title{Achieving continuously tunable critical exponents for long-range spin systems simulated with trapped ions}

\author{Fan Yang}
\affiliation{Department of Physics and Astronomy, University of British Columbia, 6224 Agricultural Road, Vancouver, BC, Canada, V6T 1Z1}
\author{Shao-Jian Jiang}
\affiliation{State Key Laboratory of Magnetic Resonance and Atomic and Molecular Physics, Wuhan
Institute of Physics and Mathematics, Chinese Academy of Sciences, Wuhan 430071,
China}
\author{Fei Zhou}
\affiliation{Department of Physics and Astronomy, University of British Columbia, 6224 Agricultural Road, Vancouver, BC, Canada, V6T 1Z1}

\date{\today}

\begin{abstract}
Quantum phase transitions are usually classified into discrete universality classes that typically only depend on symmetries and spatial dimensionalities. In this article, we demonstrate an opportunity to continuously vary the critical exponents or universalities by tuning experimental parameters in a given physical system. Particularly, we show that critical exponents in long-range spin systems simulated in ion traps can be tuned with laser detuning. We suggest that such experiments also effectively simulate some aspects of critical phenomena in conventional spin systems but in artificial non-integer spatial dimensions.
\end{abstract}


\maketitle
\section{Introduction}
Interacting quantum states including the critical ones have long been a fascinating subject.  
Various quantum systems with controllable parameters have been used to simulate quantum many-body phenomena in strongly interacting regime, especially quantum phase transitions in condensed matter physics. 
Neutral atoms in optical lattices have been applied to demonstrate phase transitions in Bose-Hubbard model \cite{Jaksch98,Greiner2002} and quantum spin models \cite{Simon2011,Greif2013}. 
In the neutral atom based simulations of quantum phase transitions, the interactions are fine-tuned to a certain strength to investigate quantum criticality.
These simulated quantum critical phenomena usually belong to a particular universality class characterized by a unique set of discrete critical exponents \cite{Sachdev}. 
As universality classes typically only depend on symmetries of interactions and spatial dimensions, they are independent of various parameters in the quantum systems used to study these phenomena.
Specifically for atoms in optical lattices \cite{Bloch2008}, critical exponents are expected to take discrete values, as a remarkable consequence of universalities. 

On the other hand, Coulomb interactions in ion traps lead to an excellent platform for quantum simulations of long-range spin models \cite{WunderlichRev,BlattRev}. 
By coupling internal states of ions to collective vibrational modes with state-dependent optical forces, 
one can achieve variable-range interactions in quantum spin models \cite{Porras2004,Porras2006,Islam2011,Britton2012}. 
The spin-spin interactions can be empirically approximated as power law with the interaction strength decaying as the distance to the power of $\sigma$ [see Eq. (\ref{xxz})], and $\sigma$ can be controlled by laser detuning $\Delta$.
This experimental platform has been applied to various quantum simulations \cite{Islam2011,Britton2012,Richerme2014,Jurcevic2014,Bohnet2016,Zhang2017}.
Here, we show that such systems can potentially be used 
to continuously tune the critical exponents in a spin model simulated in ion traps, thanks to the dependence of decay power $\sigma$ on detuning $\Delta$. We also theoretically compute 
the critical exponents as a function of detuning $\Delta$.

The tunability of decay power $\sigma$, and hence the critical exponents, allows access to different universality classes in the same physical system. 
In the following, we focus on the long-range ferromagnetic XXZ model with an external magnetic field.

\section{TUNABLITY OF THE CRITICAL EXPONENTS OF LONG-RANGE QUANTUM SPIN MODELS SIMULATED IN ION TRAPS}
The Hamiltonian of $d$-dimensional spin-$S$ ferromagnetic XXZ model is given by
\begin{equation}\label{xxz}
H=-\sum_{i\neq j}\frac{J_{ij}}2(S_i^X  S_j^X+ S_i^Y  S_j^Y+\lambda  S_i^Z  S_j^Z)-h\sum_i  S_i^Z,
\end{equation}
where $S_i^\alpha$, $\alpha=X,Y,Z$, are the spin operators at the $i$th lattice site, $J_{ij}=J_0/r_{ij}^\sigma$, $r_{ij}=|{\bf r}_i-{\bf r}_j|/R_0$ is the distance between two spins measured in lattice constant $R_0$, and $h$ is the external magnetic field. The reduced Planck constant is set to unity. 

\begin{figure}
\includegraphics[width=\columnwidth]{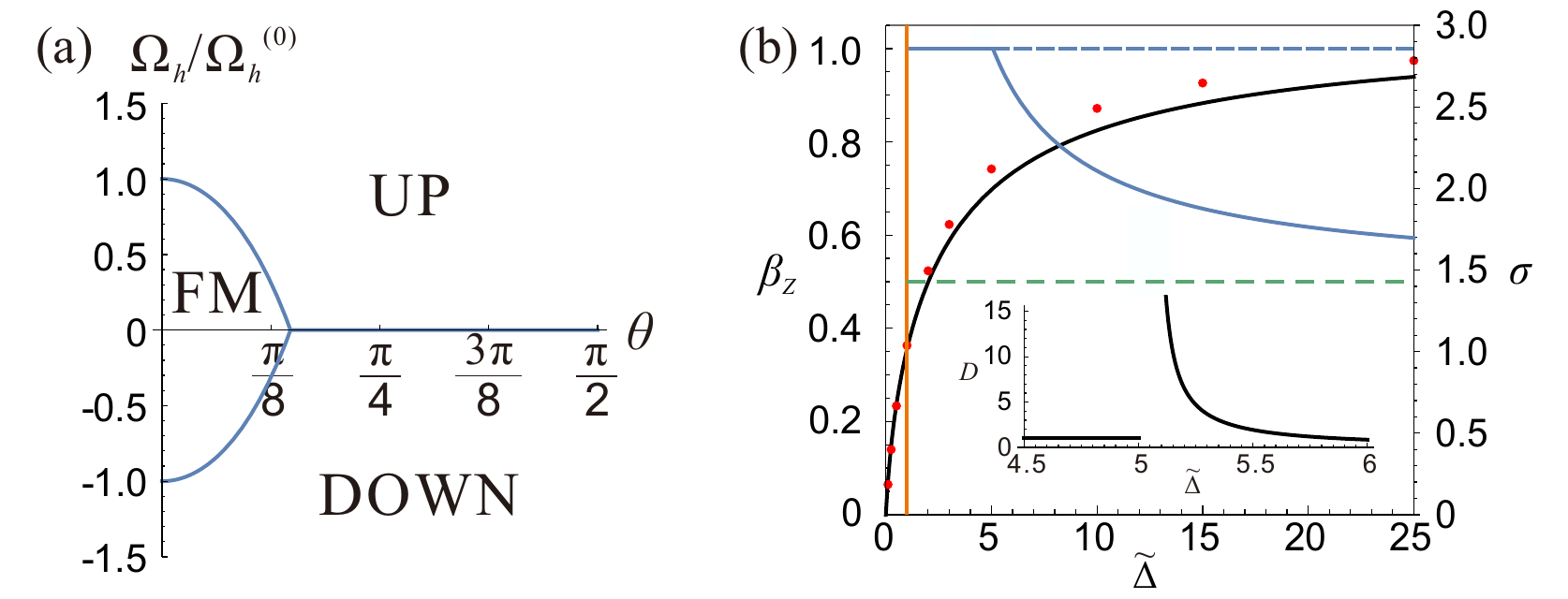}
\caption{(a) Phase diagram of the long-range XXZ model [Eq.  (\ref{xxz})] in the $\Omega_h-\theta$ plane.
$\Omega_h$ and $\theta$ are the Rabi frequency and relative phase of the resonant laser in experiments (see text).  $\Omega_h^{(0)}$ corresponds to the critical magnetic field at $\lambda=0$. In Eq.  (\ref{xxz}), $h=2\Omega_h$ and $\lambda=2\tan^2 \theta$. UP (DOWN) is the fully polarized phase along the $Z(-Z)$-direction, and FM is the XY ferromagnetic phase.
(b) Decay power $\sigma$ in the spin-spin interactions [red dots and solid black (middle) curve] defined in Eq. (\ref{xxz}) and the critical exponent ${\beta_Z}$  [solid blue (upper) curve] defined in Eq.  (\ref{magnetization}) as functions of dimensionless detuning $\tilde\Delta$ (see text) in a linear Paul trap. The orange (vertical) line labels the detuning where $\sigma=d=1$. The red dots take into account the uneven spacing between adjacent ions, while the solid black curve assumes equal spacing. The dashed blue (upper) line is the mean-field critical exponent $\beta_Z^{\text{MF}}=1$, and the dashed green (middle) line is for 1d short-range model $\beta^{1d}_Z=1/2$. Inset: $D(\sigma,d)$ as a function of $\tilde\Delta$ for $\lambda=0.5$. For $\epsilon=\phi-d<0$, the phase transition is of the mean-field type and $D=1$. For $0<\epsilon\ll1$, $D\sim\epsilon^{-1}$ as $\epsilon$ becomes small \cite{Window}.
 \label{model}}
\end{figure}

We first find the phase diagram (Fig. \ref{model}) for $\sigma>d$ where thermodynamics is well-defined.
For $0<\lambda<1$, there exists a critical magnetic field $h_c$ and three distinct quantum phases. 
When $h>|h_c|$ or $h<-|h_c|$, all the spins are fully polarized along or against the $Z$-direction, respectively; when $-|h_c|<h<|h_c|$, the spins are ferromagnetically ordered,
spontaneously breaking the $SO(2)$ symmetry in the XY-plane \cite{1d, Frerot2017, Maghrebi2017, Defenu2017}. 
The phase transition between the XY ferromagnetic (FM) phase and the fully polarized phases is continuous. 
For $\lambda>1$, the FM phase is absent and the phase transition between the two fully polarized phases is first order.

In particular, one can simulate the one-dimensional (1d) spin-$1/2$ model in Eq. (\ref{xxz}) using trapped ions \cite{Porras2004,Islam2011}.
Ions each of mass $M$ are confined in a linear Paul trap and addressed with global non-copropagating Raman beams that form a beatnote with wavevector $\delta k$ along the longitudinal direction of the ion chain.  
The trapping frequencies satisfy $\omega_x=\omega_y\gg\omega_z$ so that the ions align along the $z$-direction. 
To simplify the picture, one can map the Raman scheme to an effective two-level system with energy splitting $\omega$, corresponding to spin-$1/2$ states. 
Coulomb repulsions between ions result in collective phonon modes. 
To generate ferromagnetic interactions, a pair of in-phase off-resonant laser beams (beatnotes) with detunings $\pm\Delta$ and each of Rabi frequency $\Omega/2$ couple the internal ion states with longitudinal phonon modes, resulting in interactions analogous to the M{\o}lmer-S{\o}rensen gate \cite{Molmer1999}. 
Another resonant beam with Rabi frequency $\Omega'+\Omega_h$ and relative phase $\theta$ generates an effective magnetic field. 

In the Lamb-Dicke regime and rotating wave approximation (RWA), the effective Hamiltonian in the interaction picture with respect to the bare atomic states and longitudinal phonons is of the Ising type \cite{Islam2011}
\begin{equation}
H_{\text{Ising}}=\sum_{i\neq j}\frac{J_{ij}^{\text{eff}}}2\sigma_i^Y\sigma_j^Y+\sum_i(\Omega'+\Omega_h)({\sigma}_i^X\cos\theta-{\sigma}_i^Y\sin\theta),
\end{equation}
where $\sigma_i^\alpha$ are the Pauli operators of the $i$th ion.
The effective spin-spin interactions can be approximated as power law when $\Delta<\omega_z$ \cite{Porras2004,Britton2012,Islam2011},
\begin{equation}\label{Jij}
J_{ij}^{\text{eff}}=\Omega^2\frac{(\delta k)^2}{2M}\sum_l\frac{f_{i,l}f_{j,l}}{\Delta^2-\omega_l^2}\approx-\frac{J_0^{\text{eff}}}{r_{ij}^\sigma}.
\end{equation}
Here $f_{i,l}$ is the normal mode transformation matrix, and $\omega_l$ is the frequency of the $l$th longitudinal phonon mode. The RWA requires $\eta_{i,l}\Omega,\Omega',\Omega_h\ll|\Delta-\omega_l|$, where $\eta_{i,l}=\delta kf_{i,l}/\sqrt{2M\omega_l}$  are the Lamb-Dicke parameters. 
$\sigma$ can be continuously varied between $0$ and $3$ with laser detuning $\Delta$ \cite{Islam2011,Britton2012}.
To generate XXZ interactions, we move to another interaction picture with respect to $H_0'=\Omega'\sum_i({\sigma}_i^X\cos\theta-{\sigma}_i^Y\sin\theta)$ \cite{Cohen2014,Cohen2015}.
In the RWA, where $J_0^{\text{eff}}\ll\Omega'$, we find the effective Hamiltonian in the rotated basis $\sigma^X\to(-{\sigma'}^Y\sin\theta-{\sigma'}^Z\cos\theta)$,  $\sigma^Y\to(-{\sigma'}^Y\cos\theta+{\sigma'}^Z\sin\theta)$ and $\sigma^Z\to-{\sigma'}^X$,
\begin{multline}\label{Heff}
H_{\text{eff}}=-\sum_{i\neq j}\frac{\left|J_{ij}^{\text{eff}}\right|}2\Big[({\sigma'_i}^X{\sigma'_j}^X+{\sigma'_i}^Y{\sigma'_j}^Y)\frac{\cos^2\theta}2\\
+{\sigma'_i}^Z{\sigma'_j}^Z\sin^2\theta\Big]-\Omega_h\sum_i{\sigma'_i}^Z.
\end{multline}
Comparing this with Eq.  (\ref{xxz}), we have $J_{ij}=2J_{ij}^{\text{eff}}\cos^2\theta$, $h=2\Omega_h$ and $\lambda=2\tan^2\theta$.
 
Below we demonstrate that this experimental scheme can be used to measure the continuously varying critical exponents controlled by laser detuning.
Near the critical field, the $Z$-component magnetization in the FM phase  can be shown to have the scaling form
\begin{equation}\label{magnetization}
m_Z(h) =S\left(1- D(\sigma, d) \left|\frac{h_c-h}{h_c}\right|^{{\beta_Z(\sigma, d)}}\right).
\end{equation}
Here ${\beta_Z}=d/\phi$ for $\phi>d$ and ${\beta_Z}=1$ for $\phi\leq d$ where $\phi=\min(2,\sigma-d)$. 
$D$ is a dimensionless prefactor associated with magnetization when $h$ is varied near its critical value $h_c$. We find $D$ to be a function of $\sigma$  and $d$ (see below and Fig. \ref{model}). $S$ is the total spin at each lattice site as defined in Eq. (\ref{xxz}). Therefore, both $D$ and ${\beta_Z}$ can be continuously varied by laser detuning $\Delta$ in experiments.

Note that simple mean-field analysis yields a constant exponent $\beta_Z^{\text{MF}}=1$, and for short-range XXZ model in 1d, $\beta^{1d}_Z=1/2$ \cite{Zhou2004}.
In the conventional paradigm, mean-field transitions are most relevant above the upper critical dimension ($d=2$ for the short-range XXZ model); below the upper critical dimension, one expects non-mean-field behaviors.
For the long-range model under consideration, the mean-field physics becomes relevant even in 1d because of the dependence of $\sigma$ on the laser detuning. See more discussions after Eq.  (\ref{RGE}) as well as in Fig. \ref{model}.

We illustrate $\sigma$, $\beta_Z$ and $D$ as functions of laser detuning $\Delta$ in Fig. \ref{model} using 10 ions confined in a linear Paul trap. The ions can be approximated to be equidistant.
We define the dimensionless detuning from the center-of-mass phonon frequency $\omega_z$ as $\tilde\Delta=(\omega_z^2-\Delta^2)/\omega_c^2$. Here $\omega_c^2=e^2/(4\pi\epsilon_0R_0^3M)$ is the characteristic strength of Coulomb interactions.
$\sigma$ varies continuously between $0$ and $3$ with dimensionless detuning $\tilde\Delta$. As a result, the critical exponent ${\beta_Z}$ also varies continuously with $\tilde\Delta$.

\begin{figure}
\includegraphics[width=\columnwidth]{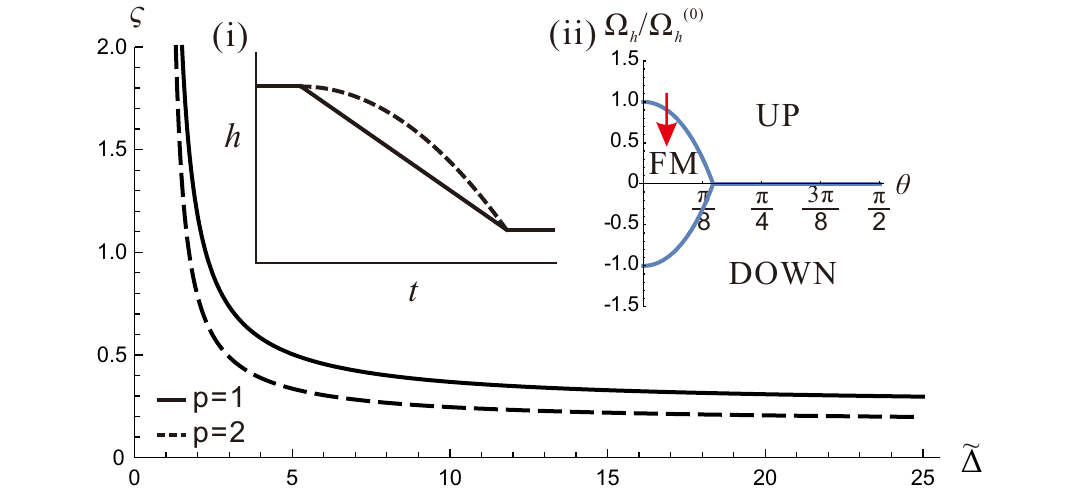}
\caption{$\zeta$, the exponent of domain wall density in a quench experiment as a function of  dimensionless detuning $\tilde\Delta$. The magnetic field varies as $h(t)=h_0-\upsilon t^p$. Inset (i) shows the ramping of the effective magnetic field $h$ controlled by the amplitude of the resonant laser. The arrow in inset (ii) shows the quench direction.\label{dynamics}}
\label{quench}
\end{figure}

Furthermore, one can study the non-equilibrium quantum spin dynamics when the external magnetic field is quenched across the critical line as $h(t)=h_0-\upsilon t^p$, $\upsilon>0$.
This quench effectively simulates the Kibble-Zurek mechanism of topological defect creation in ion traps \cite{Kibble76,Zurek85,Zurek2005}. 
XY ferromagnetic domains are expected to emerge in this process with characteristic length scale $\mathcal L^*\sim\upsilon^{-\nu/(z\nu p+1)}$ \cite{Polkovnikov2011}. Here $z$ is the dynamical critical exponent \cite{Sachdev} and $\nu$ is the critical exponent of correlation length $\xi$ defined by $\xi\sim|h-h_c|^{-\nu}$. 
In our model, $\nu=1/\phi$ and $z=\phi$ (see below). 
One can experimentally observe the scaling of domain wall density $\rho\sim\upsilon^\zeta$ during the quench process, where $\zeta={d/[\phi(p+1)]}$. 
The scaling exponent $\zeta$ also continuously depends on laser detuning $\Delta$ (Fig. \ref{dynamics}). 

To experimentally observe the continuously tunable critical exponents, one can start with $h_0>|h_c|$ and initialize all the spins along the magnetic field (see Fig.\ref{quench}). 
For each given detuning, one ramps down the magnetic field (adiabatically for $\beta_Z$ and non-adiabatically for $\zeta$). 
Noticing that the effective Hamiltonian is in the second interaction picture, one needs to follow the procedures described in Ref. \cite{Cohen2015} to measure the magnetization.

It is experimentally beneficial to couple the internal ion states with transverse phonon modes \cite{Islam2011}, 
where the power-law interactions can be realized for $\Delta>\omega_x(\omega_y)$;
however, these interactions are antiferromagnetic. 
In adiabatic simulations, one can use the antiferromagnetic Hamiltonian $H_{\text{AFM}}=-H_{\text{FM}}$ to study ferromagnetic models by following the highest excited state of $H_{\text{AFM}}$, effectively flipping the sign of the Hamiltonian \cite{Islam2011}.
For a large number of ions in the linear Paul trap, the equidistance approximation is no longer valid, and the nearest-neighbor spin-spin interactions become inhomogeneous. As a result, the critical magnetic field varies from site to site. Fortunately, linear Paul traps can be modified to achieve equidistant ion chains \cite{Johanning2016}.

Although we have focused on linear ion traps, the general scaling analysis of the long-range spin model applies to arbitrary values of $\sigma$ and $d$ for $\sigma>d$. However, for $d\geq 2$, we always have $\phi\leq d$, and $\beta_Z=1$ is fixed at the mean-field value while scaling exponents such as $\nu$, $z$ and $\zeta$ are still tunable.

\section{THEORETICAL COMPUTATION OF THE CRITICAL EXPONENTS AND THE RELATION TO FRACTIONAL DIMENSIONS}
At last, we theoretically compute the critical exponents of the long-range XXZ model.
The key idea is to reformulate the original model in a particle representation of interacting spins.
We employ Holstein-Primakoff transformation to map the long-range ferromagnetic spin system to an interacting Bose gas of magnons \cite{Zhou2004,Jiang2016}.
Unlike the short-range spin models where low-energy magnons display approximate Galilean 
invariance, the long-range spin-spin interactions introduce long-range hopping of magnons, which can break the Galilean invariance even in the limit of $k \rightarrow 0$ \cite{Jiang2016}. 
Consequently, the magnon dispersion is anomalous with $\mathcal{E}_k \sim k^\phi$ [see Eq.  (\ref{hm})]. This takes into account the most important aspect of the long-rang interactions in the original
spin model. 
Furthermore, long-range interactions between magnons are strongly renormalized to effective {\em short-range} ones, a limit we focus on below \cite{irrelevant}.

Let $a_k^\dagger$ ($a_k$) be the creation (annihilation) operator of magnons with momentum $k$. Near the critical point, where the average number of magnons per site is much less than $2S$,
one finds an effective particle representation of the long-range XXZ model in $d$-dimensional hypercubic lattice \cite{irrelevant,BoseHubbard},
\begin{equation}\label{hm}
H_m\approx\sum_k\mathcal E_ka_k^\dagger a_k+\frac {g_2  }{2\mathcal V}\sum_{k_1,k_2,k_3}a_{k_1}^\dagger a_{k_2}^\dagger a_{k_3}a_{k_1+k_2-k_3},
\end{equation}
where we rescale the Hamiltonian such that $\mathcal E_k=\frac12{k^\phi}-\mu$.
$\mu=h'_c-h'\propto h_c-h$ is the chemical potential, $h_c=S\mathcal V^{-1} R_0^d(1-\lambda)\sum_{i\neq j} J_{ij}$, $h'_c$ and $h'$ are the rescaled critical and external magnetic fields, $\mathcal V$ is the volume of the system and $g_2  =h'_cR_0^d/S$ is the rescaled two-body interaction constant. For $0<\lambda<1$, the magnon interactions are repulsive with $g_2  >0$. $n$ is the magnon density.
In this representation, the average number of magnons per site $nR_0^d$ is equal to $S-m_z$ defined in Eq.  (\ref{magnetization}).

\begin{figure}
\includegraphics[width=\columnwidth]{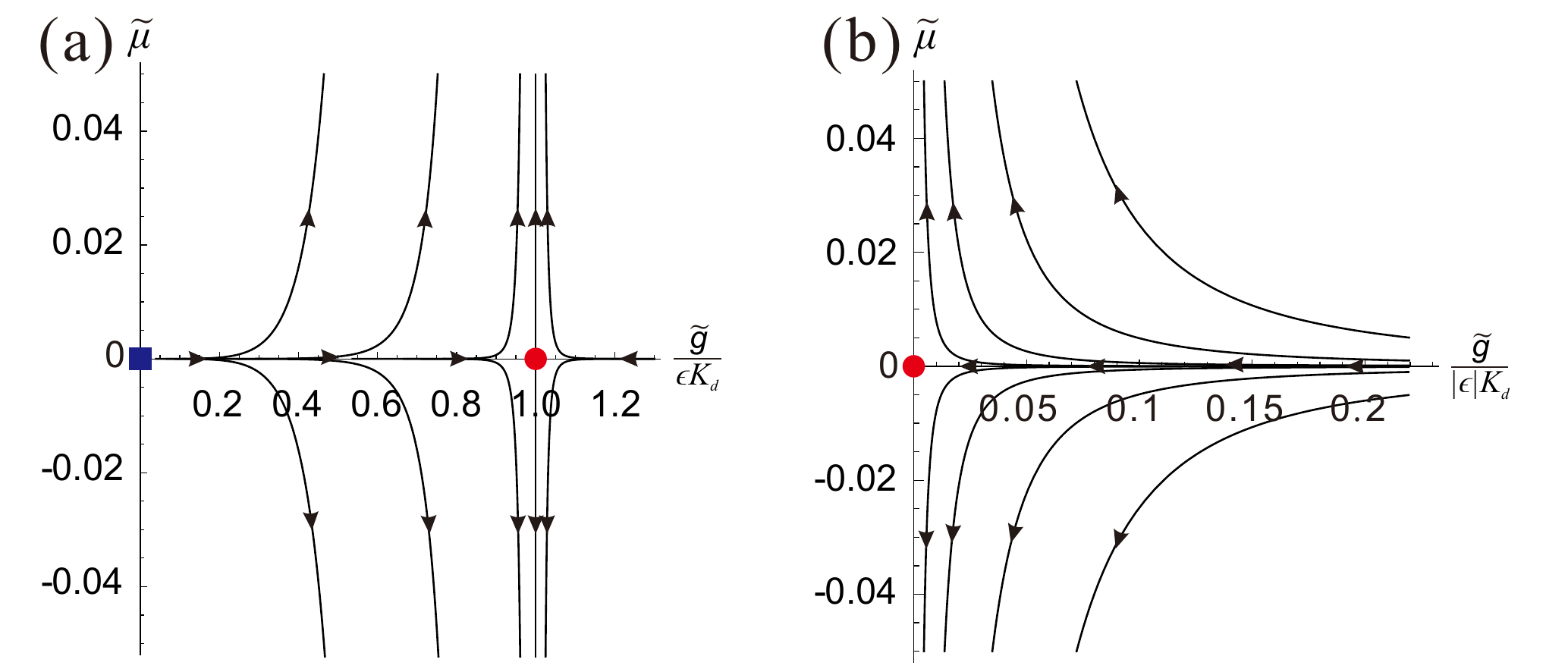}
\caption{RG flows in the plane of $({\tilde g}, \tilde{\mu})$ generated by scale transformation $\Lambda\to\Lambda e^b$, $b<0$. Arrows indicate the direction to the infrared scale.
Plotted are the solutions to Eq.  (\ref{RGE}) for different initial values. The red dots and the blue square are the stable and unstable fixed points along $\tilde\mu=0$, respectively. (a) RG flows for $\sigma=2.3$ or $\phi=1.3$ (d=1) read out at laser detuning $\tilde{\Delta}\approx 8$ in Fig. \ref{model}. (b) RG flows for $\sigma=1.7$ 
or $\phi=0.7$ at detuning $\tilde{\Delta}\approx 3$.
 \label{RG}}
\end{figure}

We compute the critical exponent $\beta_Z$ by studying the density of the interacting magnons in the presence of external field $h$.
As the phase transition is continuous, the magnon density increases from zero to a small finite value when $h$ decreases from the critical field $h_c$ (Fig. \ref{dynamics}).
We first obtain the critical exponent $\beta_Z$ by applying scaling analysis supplemented with renormalization group (RG) equations. 
Based on the general dimensional analysis, we can express the magnon density as a function of dimensionless coupling constant $\tilde g(\Lambda)$ and $\tilde{\mu}(\Lambda)$ via
 $n=\Lambda^df(\tilde g,\tilde\mu)$. Here $\Lambda$ is the momentum cutoff, $f$ is a scaling function, $\tilde g(\Lambda)=g_2  (\Lambda)\Lambda^{-\epsilon}$, $\tilde\mu(\Lambda)=\mu(\Lambda)\Lambda^{-\phi}$, and $\epsilon=\phi-d$.
The scaling behaviors of $\tilde g (b)$ and $\tilde \mu(b)$ under scale transformation $\Lambda \rightarrow \Lambda e^{b}$ can be obtained via RG equations, a general approach to critical phenomena \cite{WilsonRev,Fisher72,Zhou2012,Wu2017}. 
Following the standard calculations, we derive the following RG equations
\begin{align}
\frac{d\tilde g}{db}&=-\epsilon\tilde g+K_d^{-1}\tilde g^2,& \frac{d\tilde\mu}{db}&=-\phi\tilde\mu,
\label{RGE}
\end{align}
for $\tilde\mu\ll1$, where $K_d=2^{d-1}\pi^{d/2}\Gamma(d/2)$. 
The right-hand-sides of Eq.  (\ref{RGE}) are the standard $\beta$-functions.
Note that for short-range models $\phi=2$, so $\epsilon=2-d$ is an integer that only depends on the spatial dimension and cannot be varied continuously. 
Therefore, all the scaling properties determined by the $\beta$-function, i.e. the critical exponents, are discrete.
However, for the ion-trap system under consideration both $\phi(\Delta)$ and $\epsilon=\phi(\Delta)-d$ can be continuously tuned by laser detuning $\Delta$, as discussed before. 
This makes ion traps a very unique experimental platform for studying critical phenomena.
For instance, as $\epsilon$ is generically a continuous variable in the long-range spin model, the RG flows of $\tilde{g}(b)$ (Fig. \ref{RG}) effectively illustrate the corresponding scaling behaviors of conventional short-range spin models but in artificial non-integer dimensions. 
Flows of $(\tilde{g}(b), \tilde{\mu}(b))$ generated by the scale transformations 
visually represent the solutions to Eq.  (\ref{RGE}) and 
are numerically plotted in Fig. \ref{RG} for different decay powers $\sigma({\Delta})$.

Scale invariant fixed points under scale transformations are represented by the zeros of the $\beta$-functions and play pivotal roles in quantum critical phenomena.
RG flows near the fixed points dictate the scaling behaviors of $\tilde g$ and $\tilde \mu$, and hence $\beta_Z$ (Fig. \ref{RG}).
As in the standard Wilson theory of critical phenomena \cite{WilsonRev}, only the infrared stable fixed points along the direction of $\tilde{g}$ are relevant to the critical phenomena
driven by external field $h$. The infrared unstable fixed points are only relevant for the physics far away from the critical regime, which we shall not elaborate here. 

For $\epsilon <0$ or $\phi(\Delta) <d$, the only (non-negative) fixed point is the non-interacting fixed point $\tilde g^*=0$, which is infrared stable along the axis of $\tilde\mu=0$ as shown in Fig. \ref{RG}.
The dilute magnons are weakly interacting near the critical line,
and the phase transition can be described by mean-field theory with $\beta_Z=1$, {\it independent} of $\phi$ or laser detuning $\Delta$. 
For this reason, we focus on the regime $\epsilon > 0$ or $\phi({\Delta}) > d$, where the scaling behaviors strongly depend on $\phi$ and hence the laser detuning.
In this regime, the RG equations yield, in addition to the infrared {\it unstable} fixed point $\tilde g^*=0$, an infrared {\it stable} fixed point $\tilde g_c^*=\epsilon K_d$ corresponding to the strongly interacting  magnon gases. 
The properties of the quantum phase transition are dictated by $\tilde g_c^*$, analogous to the Wilson-Fisher fixed point below the upper critical dimension.
Near the critical point $(\tilde g_c^*=\epsilon K_d,\tilde\mu_c^*=0)$, the scaling properties of $\tilde{g}$ and $\tilde{\mu}$ can be obtained by linearizing the $\beta$-functions around this point. 
As a result, we have, under the scale transformation $\Lambda \rightarrow \Lambda e^b$,
\begin{equation}\label{scaling}
n=\Lambda^d e^{db} f_\epsilon (\tilde g_c^*+\delta \tilde g e^{\epsilon b},\tilde\mu e^{-\phi b}),
\end{equation}
where $\delta\tilde g$ is the deviation of $\tilde g(\Lambda)$ from $\tilde g_c^*$. $\delta\tilde g e^{\epsilon b}$ becomes vanishingly small in the long wavelength limit as $b\to -\infty$. 
Since $n$ is independent of the momentum cutoff, or $e^b$, we can set $\tilde\mu e^{-\phi b}$ to be a small constant or simply $\epsilon$ to obtain the scaling relation \cite{scale},
\begin{equation}\label{mdensity}
n \approx  \epsilon^{-d/\phi} f_\epsilon(\tilde g_c^*,\epsilon)\mu^{d/\phi}.
\end{equation}
This yields $\beta_Z=d/\phi$ in the spin model \cite{footnote}. 

The prefactor $f_\epsilon(\tilde g_c^*,\epsilon)$ can be computed quantitatively using the $\epsilon$ expansion when $\epsilon$ is small.
Evaluating $f_\epsilon$, or equivalently $D(\sigma, d)$ in Eq.  (\ref{magnetization}), is equivalent to computing the {\em  full equation of state} of dilute interacting magnons further taking into account the condensation
of magnons.  
We employ a more sophisticated loop expansion technique recently developed for strongly interacting Bose gases by Jiang {\it et al.} \cite{Jiang2014,Jiang2016}.
We separate the condensed magnons by writing $a_k^{(\dagger)}=\sqrt {\mathcal V n_0} \delta_{k,0}+a_k^{(\dagger)}(1-\delta_{k,0})$ and treat the density of the condensed magnons $n_0$ as an external classical field.
We compute the effective potential of the condensate $\Phi(\mu, n_0)$ diagrammatically and find the equation of state from the minimization condition ${\partial \Phi(\mu,n_0)}/{\partial n_0}=0$, and $n=-{\partial \Phi(\mu, n_0)}/{\partial \mu}$.
The potential $\Phi$ can be organized in terms of the number of loops involved. Contributions of diagrams with $L=2,3,...$ loops are suppressed by a factor of $\epsilon^L$ for small $\epsilon$. To the lowest order of $\epsilon$, we have $\Phi(\mu,n_0)=(1/2)n_0^2g(2\mu,0)-\mu n_0$, $\mu= n_0g(2\mu,0)$, and $n=(1+O(\epsilon))n_0$,
where $g(\omega+2\mu,k)$ is the $T$-matrix of two-body scattering with total frequency $\omega$ and momentum $k$, and $g(2\mu,0)\approx\epsilon K_d(2\mu)^{\epsilon/\phi}$. Therefore, we find the equation of state in the lowest order approximation $n=(2^{\epsilon/\phi}\epsilon K_d)^{-1}\mu^{d/\phi}$.  Comparing with Eq.  (\ref{magnetization}), one finds that 
$D(\sigma,d)\sim\epsilon^{-1}$ when $\epsilon$ is small \cite{Window}.  Detailed dependence of $D$ on dimensionless detuning $\tilde{\Delta}$ is plotted in Fig. \ref{model}.
This approach is fully controllable especially for small $\epsilon$.
Note that because of the condensate, the $U(1)$ symmetry [$SO(2)$ symmetry in the XXZ model (see FM phase in Fig. \ref{model})] is broken spontaneously even in 1d, a characteristic of long-range spin-spin interactions.
However, near $\sigma=3$ or in the short-range model, we expect the symmetry breaking state of FM phase to be replaced by a state with quasi-long-range order. The phase diagram at $h=0$ was studied in Refs. \cite{Frerot2017,Maghrebi2017,Defenu2017}.

In addition, we find that $|h_c-h|\sim\mu\sim\xi^{-\phi}$, so $\nu=1/\phi$. 
Considering that $\mathcal{E}_k =\frac12{k^\phi}+h'-h'_c$ above the critical line in Fig. \ref{model}, we directly observe the dynamical critical exponent $z=\phi$.
Below the critical line, by diagonalizing the quadratic terms using Bogoliubov transformation after separating the condensed magnons  in Eq.  (\ref{hm}), we have $\omega_k\sim (h'_c-h')^{1/2} k^{\phi/2}$ in the small $k$ limit. 
The characteristic time scale at the wavelength of  $\xi={(h'_c-h')}^{-1/\phi}$ is again given by $\xi^{-\phi}$, hence $z=\phi$.
\section{conclusions}
In this article, we have investigated an interesting magnetic system whose critical exponents can be continuously tuned and this tunability has the potential to be achieved in experiments by simulating the system with trapped ions.
This platform can be used to simulate quantum phase transitions of desired universality classes and study critical phenomena not easily accessible in  conventional quantum systems.

\section{acknowledgment}
\begin{acknowledgments}
We acknowledge the support of an NSERC discovery grant.
FZ is a fellow of Canadian Institute for Advanced Research (CIFAR).
\end{acknowledgments}


\begin{thebibliography}{99}
\bibitem{Window}  However, this singularity at small but finite $\epsilon$ can only be observed near the critical line within an exponentially small window around $h_c$.
Once $h$ moves farther away from $h_c$ out of that tiny window, $D$ takes the value of $\ln |h_c/(h-h_c)|$. When $\epsilon=0$, $D$ saturates at the value of $\ln |h_c/(h-h_c)|$ near $h_c$.
\bibitem{BoseHubbard} This Hamiltonian also describes Bose-Hubbard model with long-range hopping near the Mott-insulator-superfluid transition point, $H_{BH}=-\sum_{i,j}t_{ij}a_i^\dagger a_j+\frac U2\sum_i n_i(n_i-1)-\mu_{BH}\sum_i n_i$, where $t_{ij}=t_0/r_{ij}^\sigma$. The phase transition in Bose-Hubbard model driven by $\mu_{BH}$ is analogous to the one in XXZ model driven by external field $h$.
\bibitem{scale} The exponent below is robust and independent of the values of $\tilde{\mu} e^{-\phi b}$ we set.
\bibitem{footnote} The scaling analysis above is valid near the critical field $h_c$, or for low density magnons, when $n \ll  \epsilon^{-\phi/\epsilon} g_2  ^{d/\epsilon}$. 
Further below the critical line shown in Fig. \ref{model}, the density of magnons increases and the scaling behaviors shall be controlled by the RG flows near $\tilde g*=0$. We anticipate the mean-field scaling relation with $\beta_Z=1$ to be recovered.
\end{thebibliography}

\end{document}